\documentclass{mn2e}
\usepackage{graphicx}

\title[Source subtraction via forward modeling]
{Subtraction of point sources from interferometric radio images through an algebraic forward modeling scheme}  

\author[G. Bernardi et al.]
  {G.~Bernardi$^{1}$\thanks{E-mail: gbernardi@cfa.harvard.edu},
  D.A.~Mitchell$^{1}$, S.M.~Ord$^1$, L.J.~Greenhill$^1$, B.~Pindor$^2$, R.B.~Wayth$^3$ and   \newauthor 
 J.S.B.~Wyithe$^2$\\
  $^1$Harvard-Smithsonian Center for Astrophysics, Garden Street 60, Cambridge, MA, 02138\\
  $^2$University of Melbourne, School of Physics, Parkville 3010, Australia\\
  $^3$ICRAR/Curtin Institute of Radioastronomy, GPO Box U1987, Perth, WA6845, School of Physics, Parkville 3010, Australia}

\begin{document}

\date{Accepted xxxx. Received yyyy; in original form zzzz}

\pagerange{\pageref{firstpage}--\pageref{lastpage}} \pubyear{2009}

\maketitle

\label{firstpage}

\begin{abstract}
We present a method for subtracting point sources from interferometric radio images via forward modeling of the instrument response and involving an algebraic nonlinear minimization. The method is applied to simulated maps of the Murchison Wide-field Array but is generally useful in cases where only image data are available. 

After source subtraction, the residual maps have no statistical difference to the expected thermal noise distribution at all angular scales, indicating high effectiveness in the subtraction. 
Simulations indicate that the errors in recovering the source parameters decrease with increasing signal-to-noise ratio, which is consistent with the theoretical measurement errors. 

In applying the technique to simulated snapshot observations with the Murchison Wide-field Array, we found that all 101 sources present in the simulation were recovered with an average position error of 10~arcsec and an average flux density error of 0.15\%. This led to a dynamic range increase of approximately 3 orders of magnitude. Since all the sources were deconvolved jointly, the subtraction was not limited by source sidelobes but by thermal noise.

This technique is a promising deconvolution method for upcoming radio arrays with a huge number of elements, and a candidate for the difficult task of subtracting foreground sources from observations of the 21~cm neutral Hydrogen signal from the epoch of reionization. 
\end{abstract}


\vskip 3cm

\section{Introduction}
\label{intro:Sect}

The deconvolution of radio point sources is a problem that has been studied for several decades in radio astronomy.

When calibration errors can be neglected, the problem of subtracting point sources from deconvolved radio images ultimately reduces to a problem of fitting their positions and flux densities as accurately as the instrumental noise permits. 

The methods used to deconvolve point source sidelobes are typically based on the CLEAN algorithm (Hogbom 1974; Clark 1980). The CLEAN algorithm looks for the brigthtest pixel in the image and subtracts a fraction of the dirty beam from the image at that location, forming a residual image. The search and subtraction loop is repeated until the sidelobes are reduced below the thermal noise level. 

The model components that are found through this iterative process can be convolved with a two dimensional Gaussian and introduced back into the residual image. The best estimate of flux density and position for each source is then found by fitting a two dimensional Gaussian to the source.

The subtraction of point sources performed in this way has the known problem that the dynamic range achievable is limited by pixelization effects, i.e. by the fact that data are averaged and arranged into a regular grid. Therefore even a simple point source that does not lie at the centre of the grid cell cannot be represented by a single delta function model, but requires a potentially infinite number of components to be fully represented (Briggs \& Cornwell 1992, Perley 1999).

In presence of the visibility data, the pixelization problem can be minimized and the dynamic range improved by subtraction of sources from the ungridded visibilities (Noordam \& de Bruyn 1982; Voronkov \& Wieringa 2004) and by centering the local pixel grid on the source to be deconvolved (Cotton \& Uson 2008).

When the number of antenna elements to be correlated becomes extremely large, however, it becomes harder and harder to store the visibility data and the deconvolution has to be performed on images with, again, a limitation of the dynamic range due to pixelization effects.

This is a relevant issue for upcoming radio telescopes like the Murchison Wide-field Array (MWA, Lonsdale et al. 2009) or future instrumentation like the SKA\footnote{http://www.skatelescope.org/} since they will produce a huge number of correlated visibilities.
MWA will generate data at such a rate (approximately a PByte per day) that will be impractical to store the raw visibilities and go through the traditional selfcalibration loop, and the deconvolution of radio sources will happen in the image plane.

The deconvolution of bright point sources is also a prominent issue in the view of the detection of the epoch of reionization (EoR) through the redshifted 21~cm line emission, which is one of the main goals of the MWA.

The problem of foreground subtraction for EoR experiments has been studied by various authors in the literature (Di Matteo, Ciardi \& Miniati 2004; Morales \& Hewitt 2004; Santos, Cooray \& Knox 2005; Morales, Bowman \& Hewitt 2006; Wang et al. 2006, McQuinn et al. 2006; Gleser et al. 2008; Jeli{\'c} et al. 2008; Bowman, Morales \& Hewitt 2009;  Liu et al. 2009a; Harker et al. 2009; Liu et al 2009b; Harker et al. 2010). Most of their efforts have been devoted to demonstrations that the diffuse Galactic synchrotron radiation and the classical confusion noise due to unresolved radio sources can be subtracted if it is assumed that they are spectrally smooth and absent of calibration errors. Recent observations (Ali, Bharadwaj \& Chengalur 2008; Bernardi et al. 2009; Pen et al. 2009; Parsons et al. 2010; Bernardi et al. 2010) have started to characterize the diffuse foreground component.

All the simulations conducted so far, however, have assumed that the brightest point sources were perfectly subtracted from the data. Bowman et al. (2009) and Liu et al. (2009b) indicated that point sources should be subtracted down to a 10-100~mJy threshold in order to detect the EoR. 

Datta, Bhatnagar \& Carilli (2009) and Datta, Bowman \& Carilli (2010) studied the problem of subtraction of bright sources in the presence of calibration errors and concluded that sources brighter than 1~Jy should be subtracted with a positional precision better than 0.1~arcsec if calibration errors remain correlated over $\sim$6~hours of observation. If the errors are correlated on a shorter time length, however, they will tend to average down with time and the requirement for positional accuracy will be less stringent.

Pindor et al. (2010) developed a technique based on matched filters to subtract bright point sources in MWA images in presence of diffuse emission. They showed that the dynamic range of the residual images can be improved by a factor of $\sim$2-3 in this way.

In this paper we present a method of subtracting point sources from MWA dirty images that involves forward modeling and a nonlinear minimization scheme. Forward modeling is a general concept that can be used to extract astrophysical parameters from the data.

We applied our method to simulated MWA images to show that point sources can be deconvolved with an accuracy limited by thermal noise even without storing the visibility data.

The paper is organized as follows: in Section~\ref{forward_modeling} we present the method, in Section~\ref{various_applications} we apply the method to MWA simulated images and we conclude in Section~\ref{final_conclusions}

\section{The method}
\label{forward_modeling}

The method presented here relies on the fact that the sky emission can be {\it forward modeled}. Forward modeling is a generative model, i.e., a model that is related to the astrophysical parameters to be measured, is based on physical assumptions, and can be generated a priori, independent of the actual data. Once the forward model is determined, a minimization scheme (generally nonlinear) can be implemented to fit for the astrophysical parameters of interest.

Forward modeling has already been used in several astrophysical contexts; for example, Bailer-Jones (2010) used a forward modeling algorithm to estimate stellar parameters from optical spectra. Forward modeling finds a natural application in point source subtraction from radio images where visibility data are not accessible anymore.

In this case, the forward model does not need to be approximated by any analytical function but it is simply the synthesized beam calculated at that particular position in the sky and scaled for the source flux density. In traditional radio astronomy, the synthesized beam can be considered to remain constant throughout the whole field of view. If we consider the future arrays which will operate at low frequencies, however, the synthesized beam changes as a function of position in the map due to wide field effects and direction dependent primary beams. 
If very high dynamic range imaging is required - as it is to detect the EoR signal -, the exact synthesized beam should be computed at each location in the map without relying on any analytical approximation. For the MWA, real-time calibration data will be stored in a database and will be used to generate an accurate set of visibilities for each point source of interest. These visibilities can then be imaged and averaged in the same way that the true visibilities were imaged and averaged, resulting in a synthesized beam map for each source.

In the case of point source deconvolution, the astrophysical parameteres that have to be determined via forward modeling are the position and flux density of each point source.  

For a single point-source case, our algorithm can be described as follows. 
The image pixels are grouped into an $N$-element vector $\bf y$, where $N$ is the number of pixels in the map. Right ascension, declination and flux density of the source - i.e, the parameters to be fitted - are grouped into a three-element vector $\bf x$. The forward model ${\bf m}( {\bf x})$ is also a $N$-element vector. 

\begin{figure}
\centering
  \includegraphics[width=1.0\hsize]{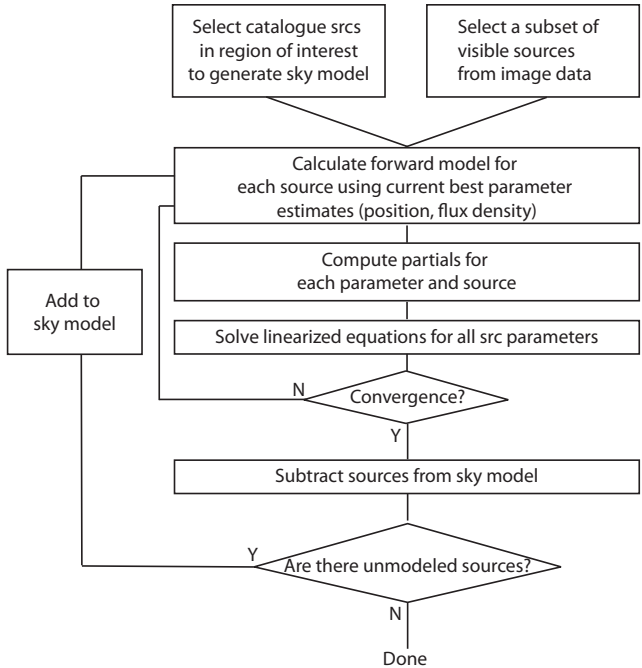}
\caption{Flow chart of the source subtraction scheme.}
\label{flow_chart}
\end{figure}

The $n$-th iteration of the method is described as follows:
\begin{enumerate}  
  \item{} generate the forward model (i.e, an image of the synthesized beam) ${\bf m}_n = {\bf m}({\bf x}_n)$, for the current parameter estimate: right ascension, declination and flux density;
  \item{} compute the $N \times 3$ Jacobian matrix, $\bf J$, which contains the derivatives of the forward model-synthesized beam with respect to the parameters computed at the current parameter estimate ${\bf x}_n$, 
\begin{eqnarray}
  \bf J_{ij} = \left( \displaystyle \frac {\partial {\bf m}_i}{\partial {\bf x}_j} \right)_{{\bf x}={\bf x}_n};	\nonumber 
\end{eqnarray}
  \item{} estimate the difference between the data and the model 
\begin{eqnarray}
  \Delta {\bf m} = {\bf y -m_n}; \nonumber
\end{eqnarray}
  \item{} estimate the shift in each parameter which is the solution of the linear system of equations 
\begin{eqnarray}
	{\bf (J^T J)} \Delta {\bf x} &=& {\bf J^T} \Delta {\bf m}	\nonumber \\
	\Delta {\bf x} &=& {\bf (J^T J)}^{-1} {\bf J^T} \Delta {\bf m};
\label{eq_minimization}
\end{eqnarray}
  \item{} compute the new estimate of the parameters 
\begin{eqnarray}
  {\bf x}_{n+1} = {\bf x}_n - \Delta {\bf x}; \nonumber
\end{eqnarray}
\end{enumerate}  
Steps (i)-(v) are repeated until convergence is reached (Figure~\ref{flow_chart}).

Equation~\ref{eq_minimization} shows that the problem of source subtraction has become a nonlinear least squares minimization. 

The forward model has a linear dependence on flux density, but nonlinear on position, therefore the partial derivatives with respect to right ascension and declination are computed numerically using finite difference approximation. 

Practically, the partial derivative with respect to right ascension is computed by generating an image of the synthesized beam with a small right ascension offset from the current estimated position. An image of the synthesized beam with a small declination offset is generated to compute the partial derivative with respect to declination. The derivative with respect to flux density is just a scaled version of the synthesized beam.

The generalization of the single point-source case to $M$ sources is straightforward, since the vector of parameters becomes a $3M$ vector, and the Jacobian matrix becomes a $N \times 3M$ matrix, and all the sources are  fitted simultaneously at each iteration. It is also important to note that the matrix ${\bf J^T J}$ that has to be inverted does not depend upon the number of pixels in the map, but only upon the number of parameters, therefore its size increases linearly only with the number of sources to be subtracted.

In principle, an initial estimate of the parameters could be obtained by generating a grid of likely models, with a range of right ascension, declination and flux densities for each source, and selecting the model, $x_0$, that best fits the data (i.e., minimizes $({\bf y}-{\bf m})^T ({\bf y}-{\bf m}))$.
In practice it is easier and faster to fit an elliptical Gaussian to the source position and use its best-fit parameters as the initial guess.

Equation~\ref{eq_minimization} can be generalized by assigning a weight to each pixel of the image. In this case it becomes the general expression for nonlinear weighted least squares:
\begin{eqnarray}
	\Delta {\bf x} = {\bf (J^T W J)}^{-1} {\bf J^T W} \Delta {\bf m},
\label{eq_minimization_w}
\end{eqnarray}
where $W$ is the $N \times N$ weight matrix. Although different weighting schemes could be explored, in the following applications of our method we will assume that $W$ is a diagonal matrix with each diagonal element equal to the signal-to-noise ratio (SNR) of the corresponding pixel.

The advantage of this method compared to other image based deconvolution techniques is that the forward model can be generated with an arbitrary level of precision in the parameter space grid and, therefore, is not affected by any pixelization effect. 
In the following section we will apply this method to simulated MWA images.

\section{Applications}
\label{various_applications}

In this section we test the method with simulated MWA images obtained through the Real Time System (RTS, Mitchell et al. 2008). The main RTS data product will be dirty images - i.e., images were the synthesized beam has not been deconvolved - integrated over a period that can range from 8~seconds to a few minutes. It is these integrated images that require subtraction of point sources. The RTS will also save calibration information (primary beam and atmospheric models) to facilitate accurate off-line deconvolution.

\subsection{Simulation setup}

We simulated a realistic MWA observation, with $20^\circ \times 20^\circ$ images covering the MWA field of view. The simulations were constructed as follows. We populated the field of view with point sources according to the following $\log N$-$\log S$ distribution:
\begin{eqnarray}
	d N = N_0 S^{-2.5} d S, \nonumber
\end{eqnarray}
where $d N$ is the differential source count, $N_0$ is the number of sources per steradian per Jy$^{-1.5}$ and $S$ is the source flux density. We have chosen $N_0$ in such a way that there are 100 sources greater than 1~Jy in a $20^\circ \times 20^\circ$ field. Random positions were assigned to the sources with no constraint on the minimum distance among them.
 
Visibility data were then created for the sources at 150~MHz, with a 2~sec cadence and over a 40~kHz channel width using the MAPS package (Wayth et al. 2010).

Random noise was added to each  $ij$ visibility, for each polarization, according to the following expression:
\begin{eqnarray}
	N_{ij} = \frac{SEFD}{\sqrt{\Delta t \Delta f}},
\label{noise_eq}
\end{eqnarray}
where we have assumed a System Equivalent Flux Density SEFD=10000~Jy, $\Delta t$=2~sec and $\Delta f$=40~kHz.

The visibilities were imaged through the RTS, which performs calibration and imaging of raw visibility data. However, since we assumed a perfect calibration in this work, only the imaging part was used. The RTS imaging pipeline is described by Ord et al. (2010) and we briefly summarize it here, referring the reader to the paper for a more detailed presentation. 

Each MWA tile is constituted of 16 dipoles arranged in a $4 \times 4$ square configuration. The dipoles are fixed to the ground, therefore their projection on the sky changes with time. The primary beam response to the sky brightness is, therefore, time variable. We have assumed that the tile primary beams can be described by the sum of 16 complex numbers that represent gain terms for the individual, known, dipole beams. Since we are not dealing with calibration, all tile beams are assumed to have the same shape, amplitude and phase. 

The RTS expects visibility data from the correlator to be integrated over 2 seconds and 40 kHz. The visibility data are then averaged and imaged over 8 seconds (Mitchell et al. 2008). Each individual 8 second snapshot is then resampled into the Healpix frame (G{\'o}rski et al. 2005) and integrated over time, with wide-field distortions corrected during the resampling. 
The time integration is perfomed, for each pixel, by summing over the measured values weighted by the complex conjugate of their primary beam response (the total weight is now the square of the beam). The sum of the weights - i.e., the square of the primary beam response integrated over the duration of the observation - is divided out at the end of the integration.

We generated images centred at 4$^h$ right ascension and $-30^\circ$ declination, which is one of the potential fields for EoR observations. We have assumed that the field was observed one hour before transit.

\begin{figure}
\centering
  \includegraphics[angle=-90,width=1.0\hsize]{32T_uv_coverage_expanded.ps}
  \includegraphics[angle=-90,width=1.0\hsize]{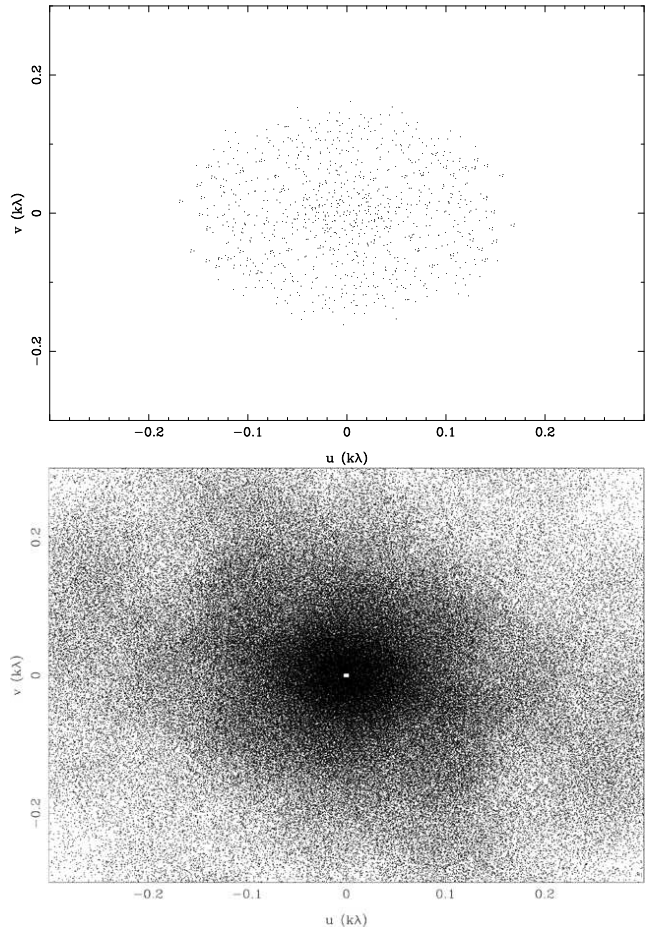}
\caption{Simulated 32T $uv$ coverage integrated over 10~minutes (top), and simulated 512T $uv$ coverage for an 8~seconds snapshot (bottom). The image centre is at 4$^h$ right ascension and $-30^\circ$ declination.}
\label{32T_512T_uv_coverage}
\end{figure}

We simulated two different sets of observations related to two different array configurations. First, we considered the 5\% protoype of the array that is currently deployed on the ground and is constituted by 32 tiles (32T). Second, we considered the full MWA configuration which will consist of 512 tiles (512T). The 32T system has less sensitivity compared to the 512T system and a coarser angular resolution since its longest baseline is $\sim$400~m whereas the longest baseline is $\sim$1500~m in the 512T configuration. The instantaneous $uv$ coverage of the 32T is also much worse than the 512T one (Figure~\ref{32T_512T_uv_coverage}).

\begin{figure}
\centering
  \includegraphics[width=1.0\hsize]{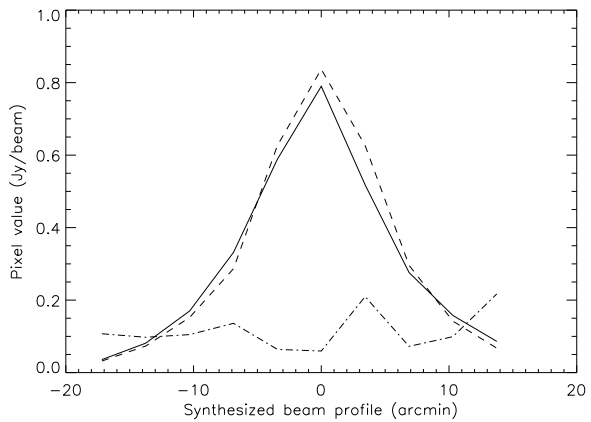}
\caption{An example of the difference between synthesized beams at two positions in the 512T simulated image. Solid line: the synthesized beam profile at the image centre. Dashed line: the synthesized beam profile 9$^\circ$ away from the image centre. Dot-dashed line: the difference between the two profiles. In this example the difference is at the 10\% level.}
\label{beam_profile}
\end{figure}

The presence of wide-field effects makes the synthesized beam position dependent even in the absence of calibration errors (Figure~\ref{beam_profile}). Since we are aiming at achieving a high dynamic range subtraction, we will generate the synthesized beam for each source at the specific source location to account for the difference.

\subsection{32T Results}

We used the 32T simulations to test the applicability of our method to integrated snapshot images. Since the long integration images taken with MWA will be obtained by co-adding individual snapshots - whose duration can vary from 8~seconds to $\sim$5~minutes - it is relevant to test the capability of the algorithm to subtract sources over co-added images, where fainter sources can become visible because of sidelobe suppression and the lowering of thermal noise.

We used a 10~min integrated image, which has an rms thermal noise of $\sim$52~mJy~beam$^{-1}$, enabling detection of sources brighter than $\sim$300~mJy. However, since the computational load increases with the number of sources and the length of the observation, we limit ourselves to the sixteen sources brighter than 4~Jy, working in a case of high SNR.

The subtraction was performed without any a priori assumption about the sky model, that is without identifying in advance sources via catalogued coordinates. This enables us to test the robusteness of the algorithm under realistic conditions (i.e., without presupposition of a sky model generated by the MWA), where sidelobe structure from sources around the sky cannot be filtered.

\begin{figure}
\centering
  \includegraphics[width=1.0\hsize]{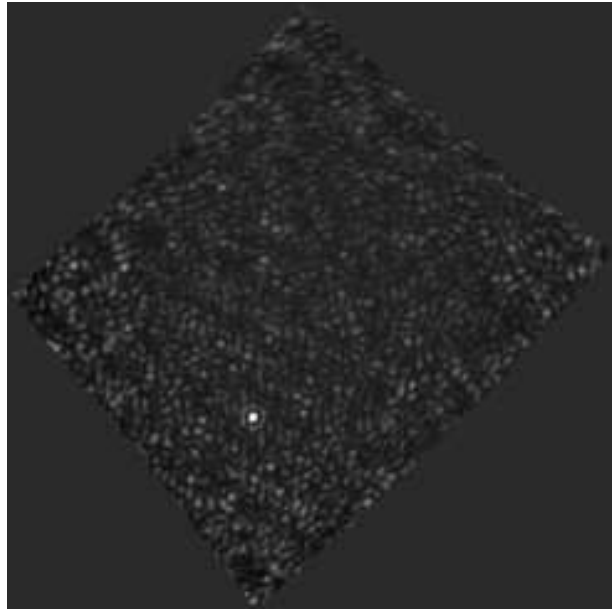}
\caption{Simulated image with the 32T $uv$ coverage integrated over 10~minutes. The black and white scale runs linearly between -10 and 50~Jy~beam$^{-1}$. The sidelobes of the dominant source ($\sim$97~Jy) obliterate the other fifteen sources between it and a 4~Jy floor.}
\label{32T_original}
\end{figure}

\begin{figure}
\centering
  \includegraphics[width=1.0\hsize]{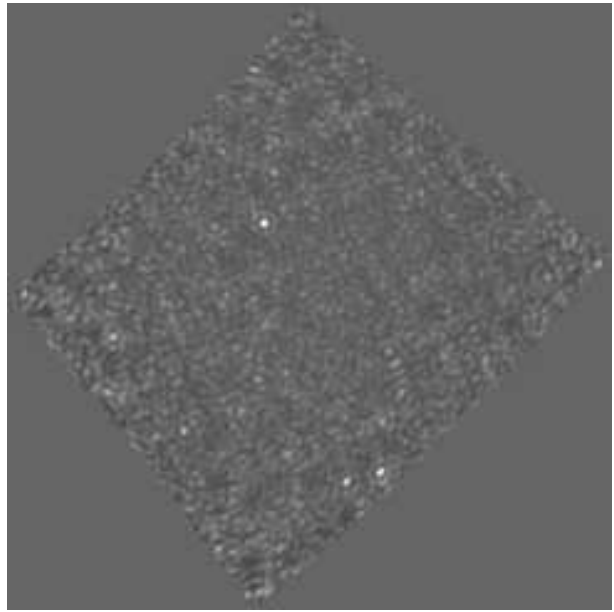}
\caption{Residual image with the brightest source subtracted, after three iterations. The black and white scale runs between -10 and 15~Jy~beam$^{-1}$. Three new sources are visible.}
\label{32T_residual1}
\end{figure}

\begin{figure}
\centering
  \includegraphics[width=1.0\hsize]{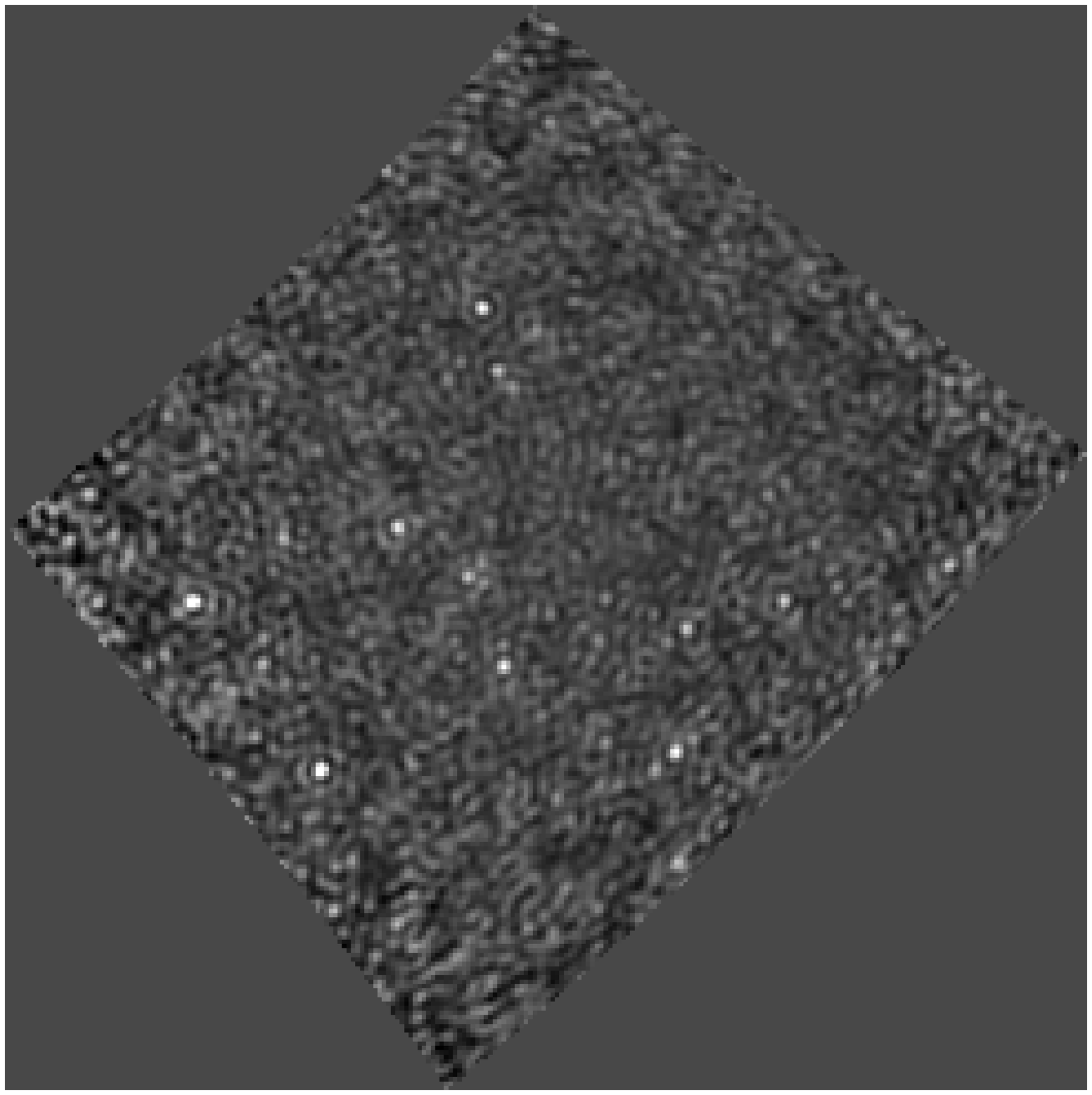}
\caption{Residual image with the four brightest sources subtracted, after three iterations. The black and white scale runs between -2 and 5~Jy~beam$^{-1}$. Seven new sources are visible.}
\label{32T_residual2}
\end{figure}

\begin{figure}
\centering
  \includegraphics[width=1.0\hsize]{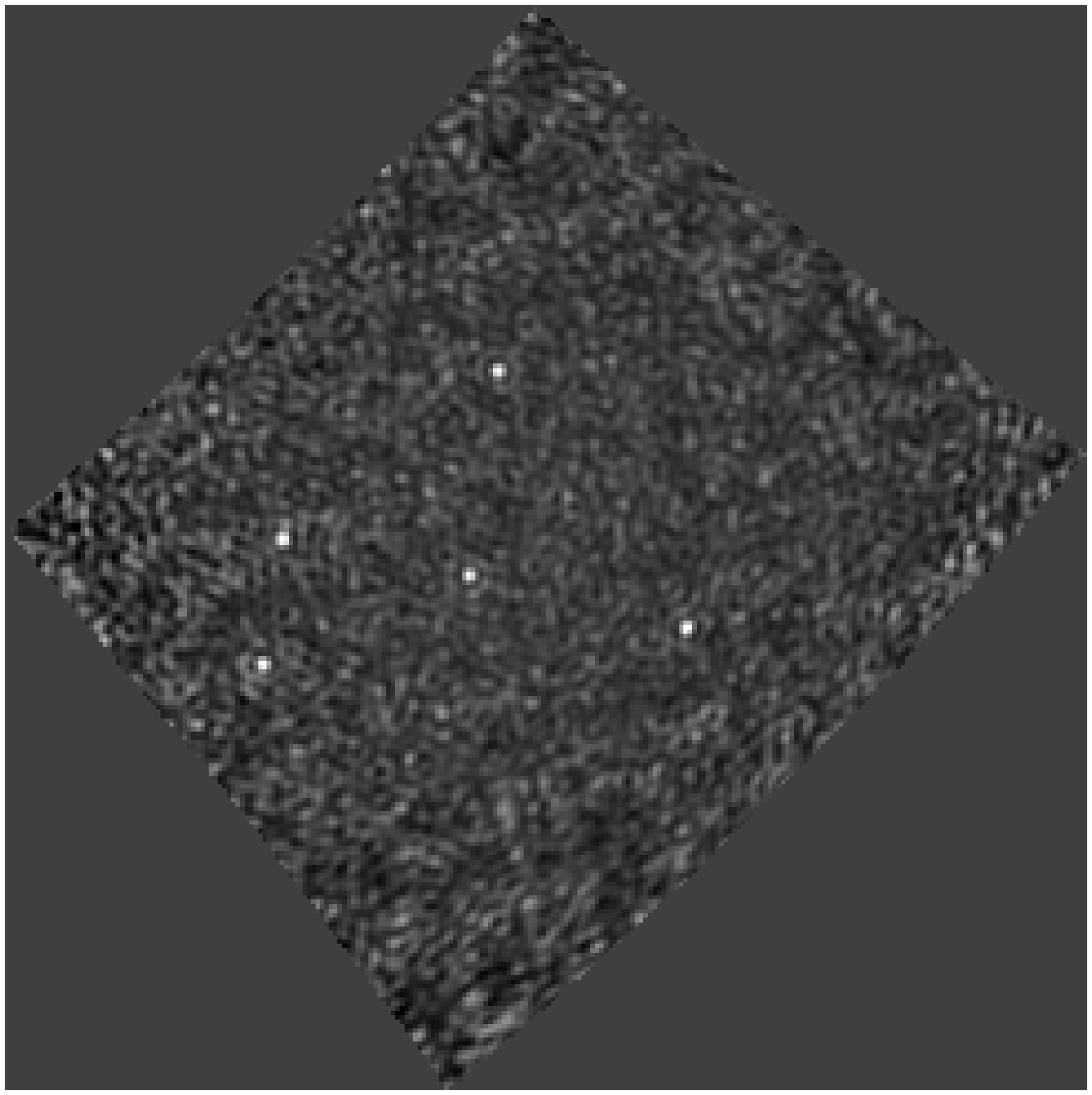}
\caption{Residual image with the eleven brightest sources subtracted, after three iterations. The black and white scale runs between -1 and 3~Jy~beam$^{-1}$. Five new sources are visible.}
\label{32T_residual3}
\end{figure}

\begin{figure}
\centering
  \includegraphics[width=1.0\hsize]{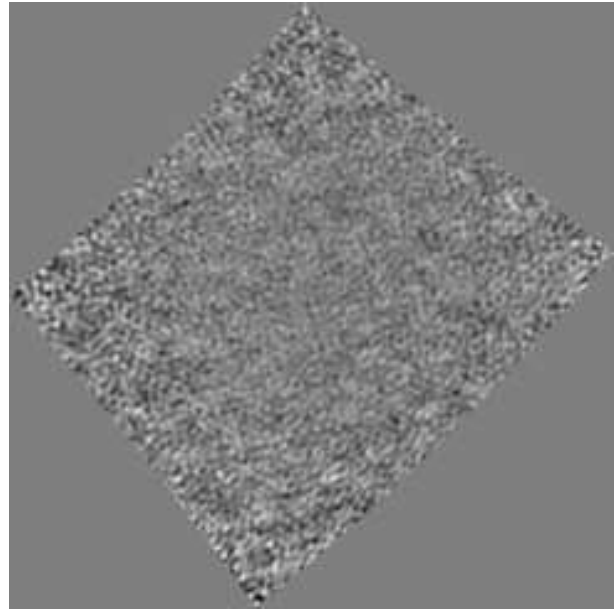}
\caption{Residual image with all the sources subtracted, after three iterations. The black and white scale runs between -0.4 and 0.4~Jy~beam$^{-1}$. The image shows only thermal noise.}
\label{32T_residual4}
\end{figure}

For a sixteen source model that includes thermal noise, only one source (97 Jy) is clearly visible because its sidelobes are bright enough to cover all the remaining sources (Figure~\ref{32T_original}). The initial guess regarding the sky brightness distribution is limited to the parameters for this one source. The subtraction was performed according to the following steps:
\begin{enumerate}  
  \item{} the first source parameters were estimated through the forward modeling minimization (three iterations, Figure~\ref{32T_residual1}); the source model was subtracted, and initial guesses obtained for the three newly visible sources;
  \item{} the four brightest sources were included in the sky model and subtracted. After three iterations another seven sources were detected in the image (Figure~\ref{32T_residual2}) and initial estimates of their parameters were made;
  \item{} a sky model made of eleven sources was subtracted. After three iterations all the remaining sources were identified (Figure~\ref{32T_residual3}) and an initial estimate of their parameters performed;
  \item{} the full sky model is minimized and subtracted jointly, giving the residual image of Figure~\ref{32T_residual4}.
\end{enumerate}  

\begin{figure}
\centering
  \includegraphics[width=1.0\hsize]{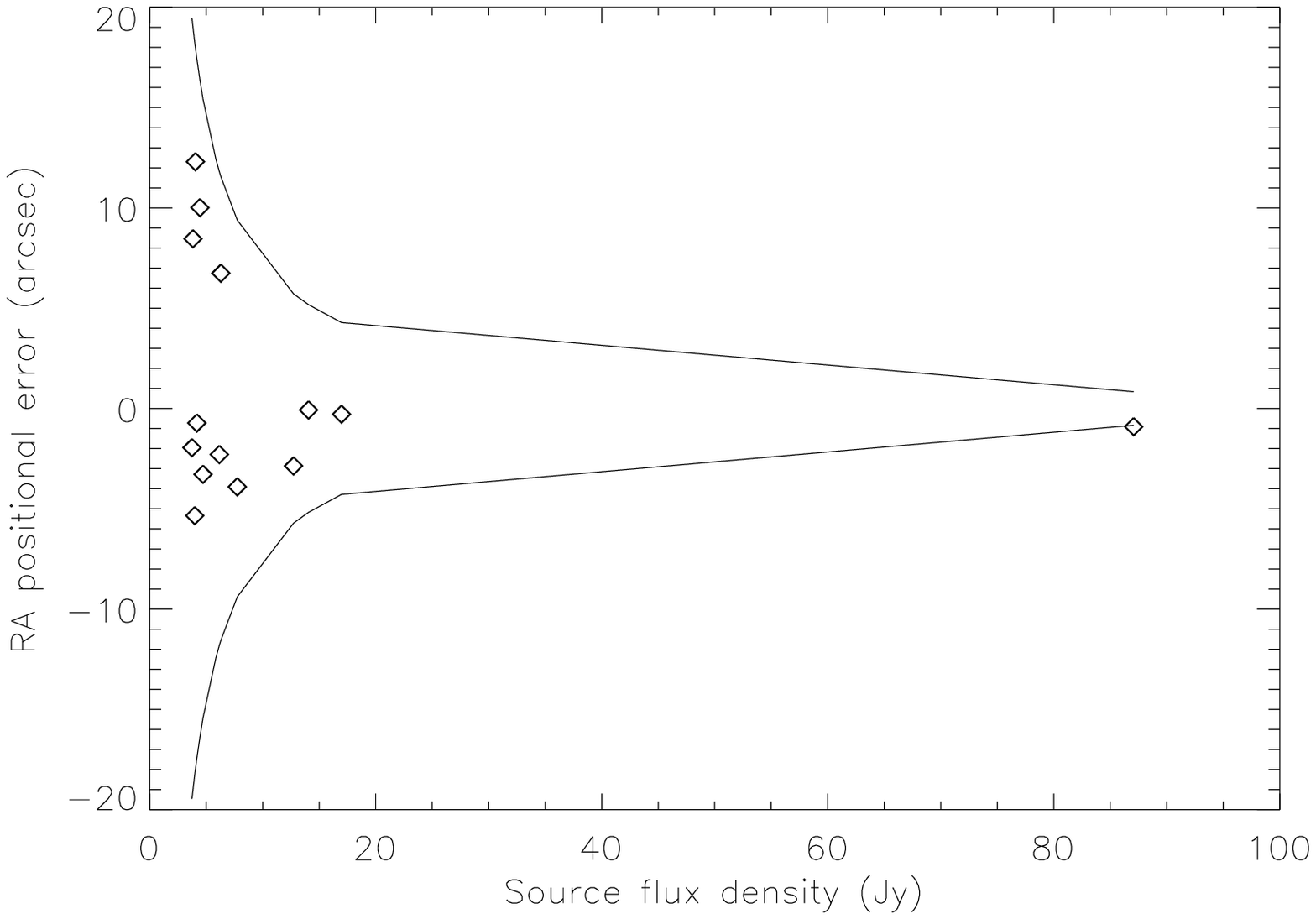}
  \includegraphics[width=1.0\hsize]{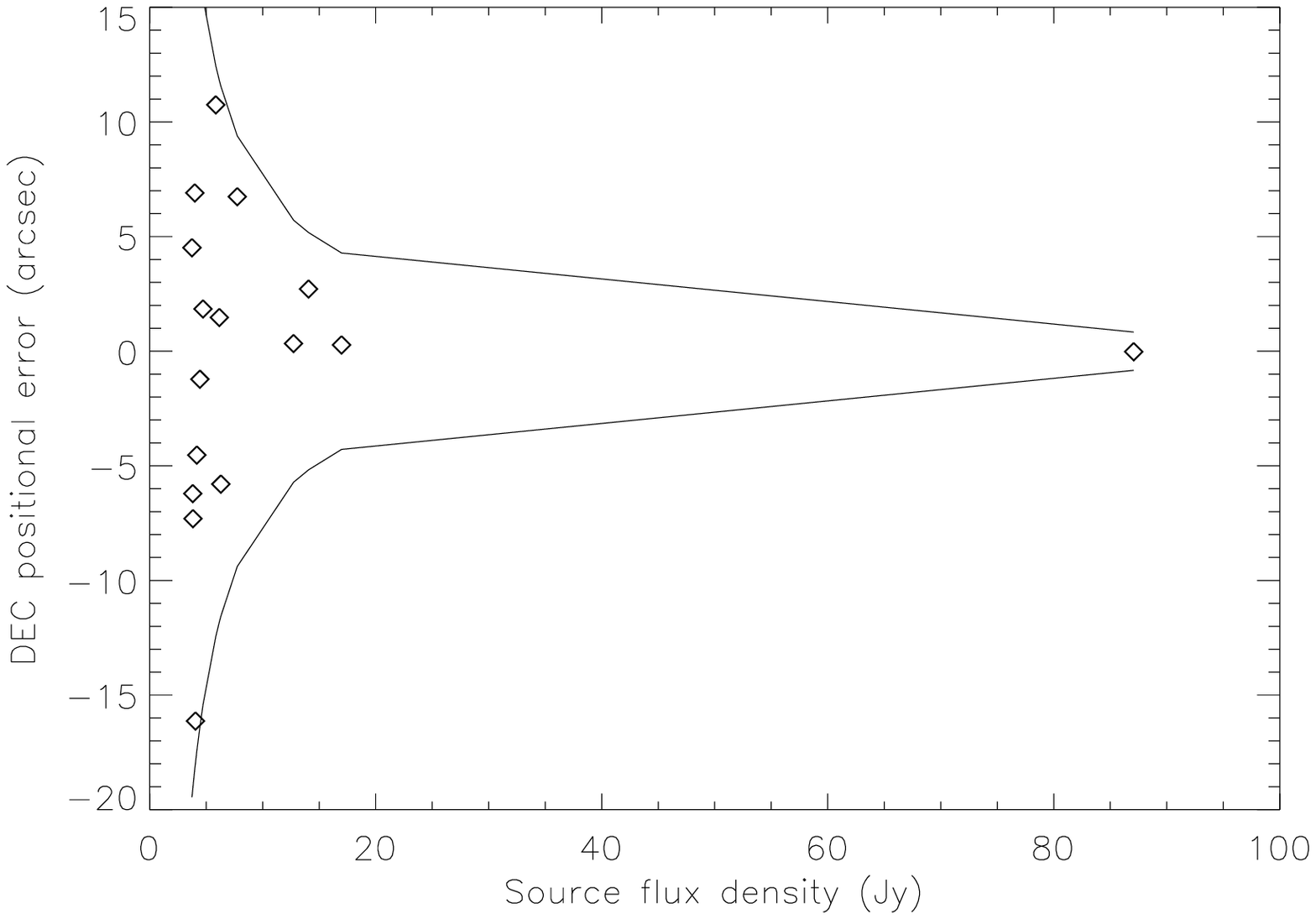}
  \includegraphics[width=1.0\hsize]{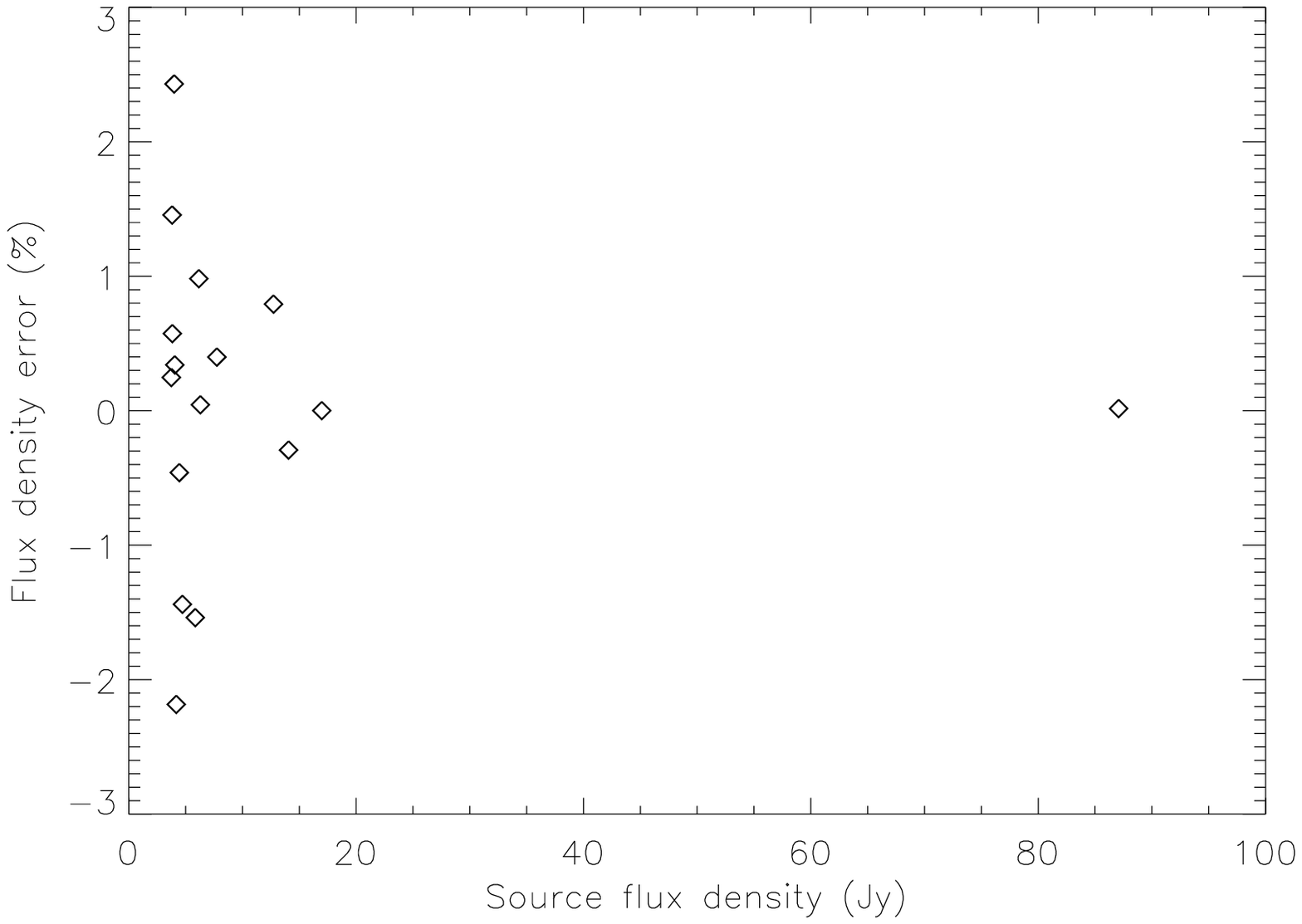}
\caption{Errors in the fitted parameters for sources in the 32T simulated image: right ascension (top), declination (middle) and flux density (bottom). Solid lines in the two upper plots indicate the envelope of the theoretical measurement errors.}
\label{err_32T}
\end{figure}

In order to characterize the statistics of the residuals and the accuracy of the subtraction, we compare the true flux densities and positions to the final estimates and to the theoretical measurement errors $\sigma^{RA,DEC}_{theor}$ computed as:
\begin{eqnarray}
  \sigma^{RA,DEC}_{theor} = \frac{\Theta_b}{2 \, \rm{SNR}} \nonumber \\
\end{eqnarray}
where $\Theta_b \sim 18$~arcmin is the synthesized beam (Figure~\ref{err_32T}). 

We observe that the error distribution narrows with increasing flux density and is within the  theoretical values. No systematic offsets appear in the recovered source parameters, based on estimates of median and rms values (Table~\ref{best_fit_32T}).
\begin{table}
 \centering 
  \caption{Median and rms values of the offset between the model and the fitted parameters for the simulated 32T image.}
  \begin{tabular}{@{}lcc@{}}
  Parameter	&	Median			&  	rms \\
  \hline
  RA		&	-0.7~arcsec		&	9.6~arcsec	\\
  DEC		&	-0.3~arcsec		&	6.6~arcsec	\\
  flux density	&	0.3\%			&	1.2\%		\\
  \end{tabular}
 \label{best_fit_32T}
\end{table}

We also computed the angular power spectrum of the residual images as (Seljak 1997, Bernardi et al. 2009):
\begin{eqnarray}
   C_\ell = \frac{\Omega}{N_\ell} \sum_{\bf l} X({\bf l}) X^*({\bf l}) 
  \label{pow_spec_def}
\label{power_spec}  
\end{eqnarray}
where $\ell=\frac{180}{\Theta}$ is the usual multipole value, $\Theta$ is the angular scale in degrees, $\Omega$ is the solid angle in radians, ${N_\ell}$ is the number of Fourier modes around a certain $\ell$ value, $X$ and $X^*$ are the Fourier transform of the image and its complex conjugate respectively and $\bf{l}$ is the two dimensional coordinate in Fourier space. The power spectrum has a bin width of $\Delta \ell = 50$. 

\begin{figure}
\centering
  \includegraphics[width=1.0\hsize]{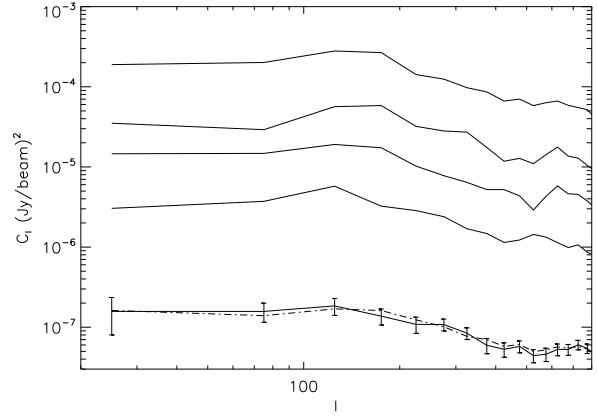}
\caption{Power spectra of residual images for an increasing number of subtracted sources. From top to bottom: after subtracting one source (Figure~\ref{32T_residual1}), after subtracting four sources (Figure~\ref{32T_residual2}), after subtracting eleven sources (Figure~\ref{32T_residual3}), after the whole sky model is subtracted (Figure~\ref{32T_residual4}). The error bars are at the 1$\sigma$ confidence level. The dashed line represents the noise power spectrum (see text for details).}
\label{power_spectrum_32T}
\end{figure}

The amplitude of the power spectrum of the residual images decreases by more than two orders of magnitude as the number of subtracted sources increases (Figure~\ref{power_spectrum_32T}).

Figure~\ref{power_spectrum_32T} also shows the noise power spectrum, estimated as the averaged power spectrum of 100 noise realizations. Each noise realization was generated by imaging visibilities which included only noise, following Equation~\ref{noise_eq}. A noise power spectrum was computed from each image. The estimated noise power spectrum was determined as the average among 100 power spectrum realizations. The error bars are the standard deviation of the 100 power spectrum realizations in each multipole bin.

It can be seen that the power spectrum of the residual image, after the full 16-source sky model is subtracted, agrees with the estimated thermal noise over the entire range of angular scales probed. This indicates that no systematic errors or statistical deviations from Gaussian distributed noise are introduced by the method and that the source subtraction is accurate down to the thermal noise level.

\subsection{512T Results}
\label{512T_Results}

\begin{figure}
\centering
  \includegraphics[width=1.0\hsize]{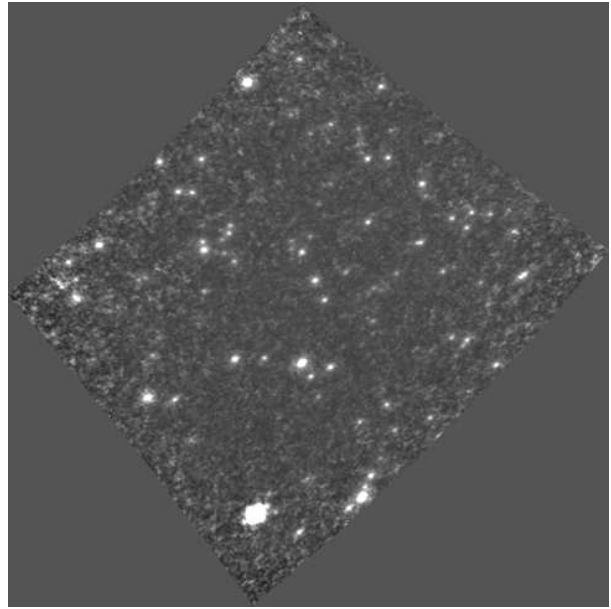}
\caption{Simulated 8~sec snapshot image with 512T $uv$ coverage. The black and white scale runs between -1 and 2~Jy~beam$^{-1}$. The very good synthesized beam has low sidelobe levels and makes most of the sources directly visibile without any subtraction.}
\label{512T_original}
\end{figure}

The 512T simulation included 101 sources brighter than 1~Jy, observed in an 8~second snapshot and in a 40~kHz channel (Figure~\ref{512T_original}). The thermal noise in the 512T image is $\sim$26~mJy~beam$^{-1}$. Unlike the 32T case, the 512T image prior to forward modeling already exhibits a great number of sources, due to the reduced sidelobes of the synthesized beam.

\begin{figure}
\centering
  \includegraphics[width=1.0\hsize]{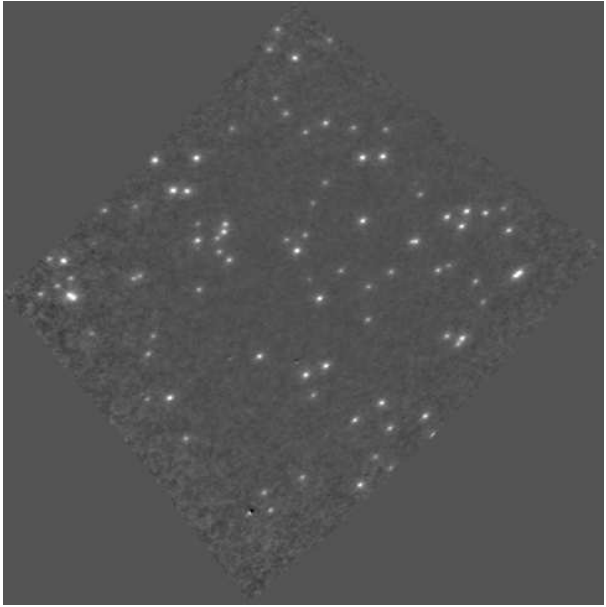}
\caption{Residual image with the brightest fifteen sources subtracted, after five iterations. The black and white scale runs between -1 and 2~Jy~beam$^{-1}$.}
\label{512T_residual1}
\end{figure}

\begin{figure}
\centering
  \includegraphics[width=1.0\hsize]{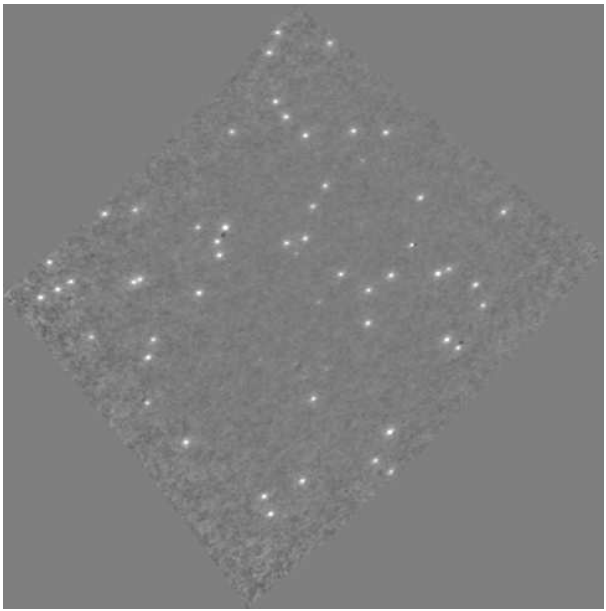}
\caption{Residual image with the 50 brightest sources subtracted, after five iterations. The black and white scale runs between -1 and 2~Jy~beam$^{-1}$. The presence of negative peaks is due to a subtraction in the absence of a full sky model.}
\label{512T_residual2}
\end{figure}

\begin{figure}
\centering
  \includegraphics[width=1.0\hsize]{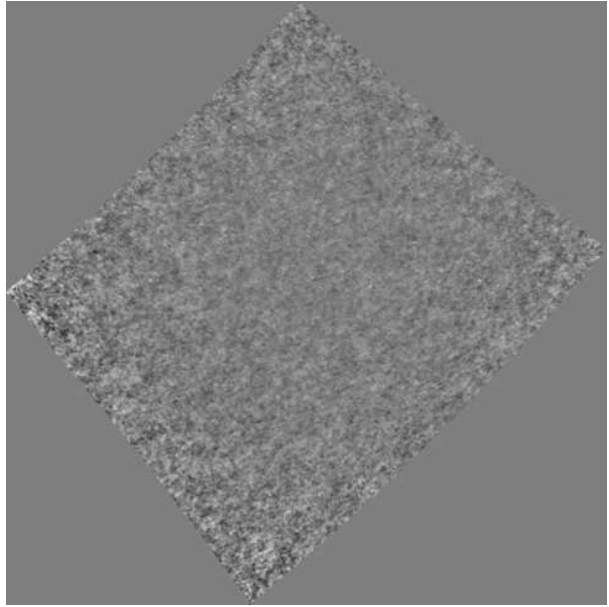}
\caption{Residual image with all sources subtracted, after five iterations. The black and white scale runs between -0.3 and 0.3~Jy~beam$^{-1}$. Only thermal noise is left after the subtraction of the whole sky model.}
\label{512T_residual3}
\end{figure}

The subtraction was performed according to the following steps:
\begin{enumerate}  
  \item{} the brightest fifteen sources were identified and an initial guess of their parameters estimated. They were then subtracted out through the minimization scheme (Figure~\ref{512T_residual1});
   \item{} another 35 sources were identified and their parameters estimated. The joint fit is now performed on 50 sources simultaneously  (Figure~\ref{512T_residual2});
   \item{} all the sources were included in the sky model. The minimization was carried out for all the 101 simultaneously and convergence was reached after five iterations (Figure~\ref{512T_residual3});
\end{enumerate}  

The final residual image after the whole sky model is subtracted is consistent with the initial thermal noise level, indicating an accurate subtraction of the sources. It is worth noticing that in the intermediate steps, when only a partial sky model is subtracted, residual features due to an imperfect subtraction exist, and appear as positive adjacent to negative peaks.

\begin{figure}
\centering
  \includegraphics[width=1.03\hsize]{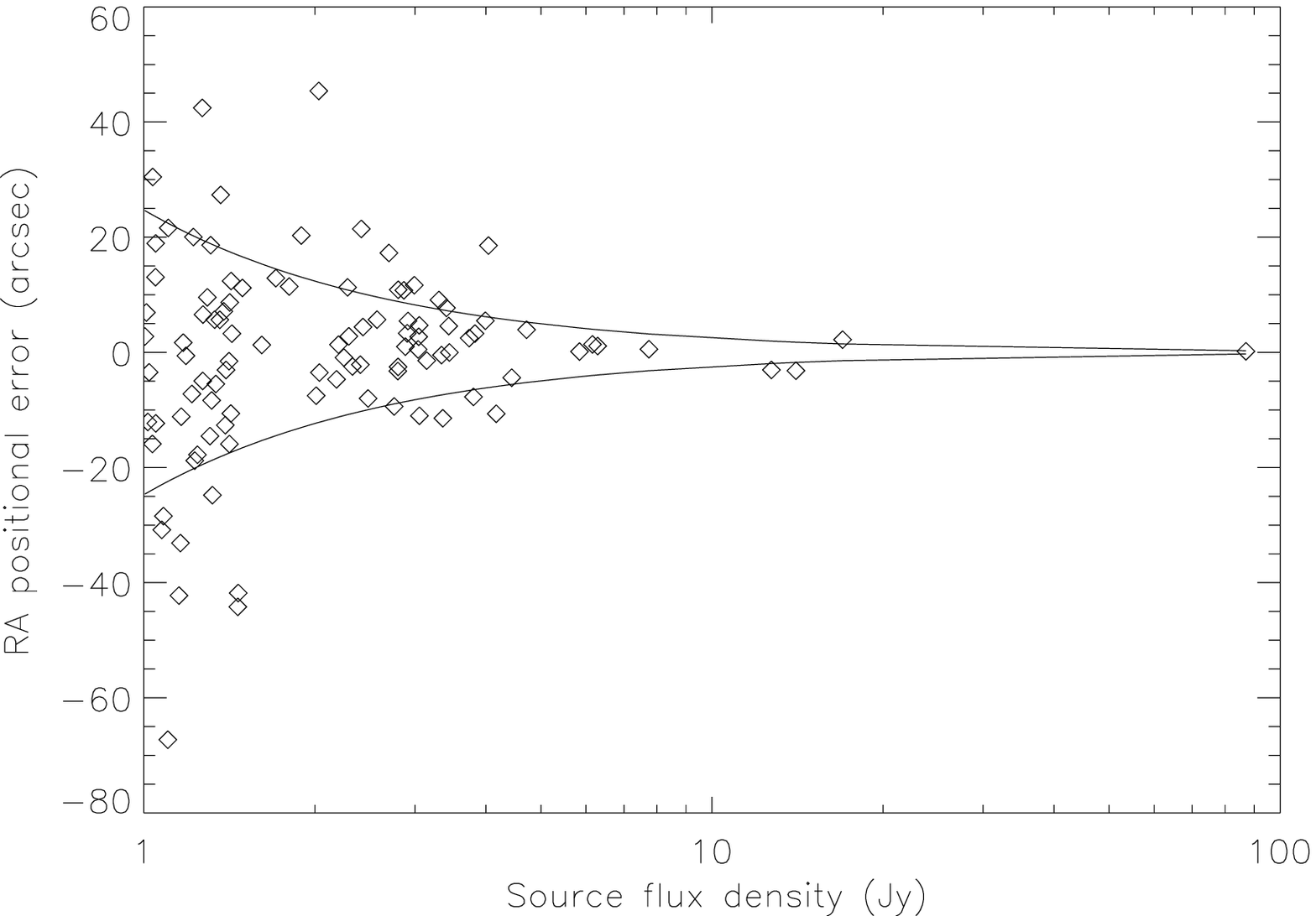}
  \includegraphics[width=1.03\hsize]{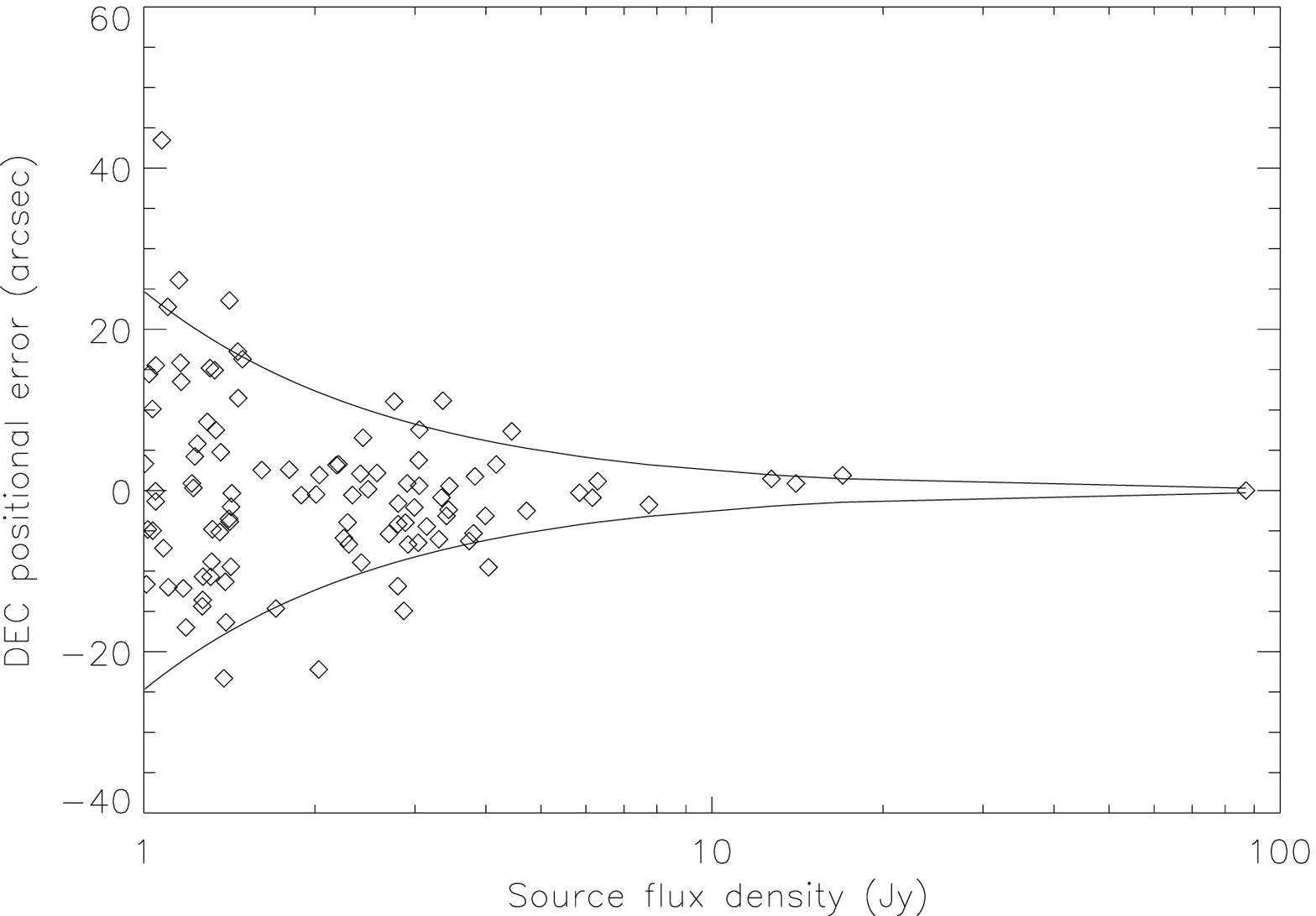}
  \includegraphics[width=1.03\hsize]{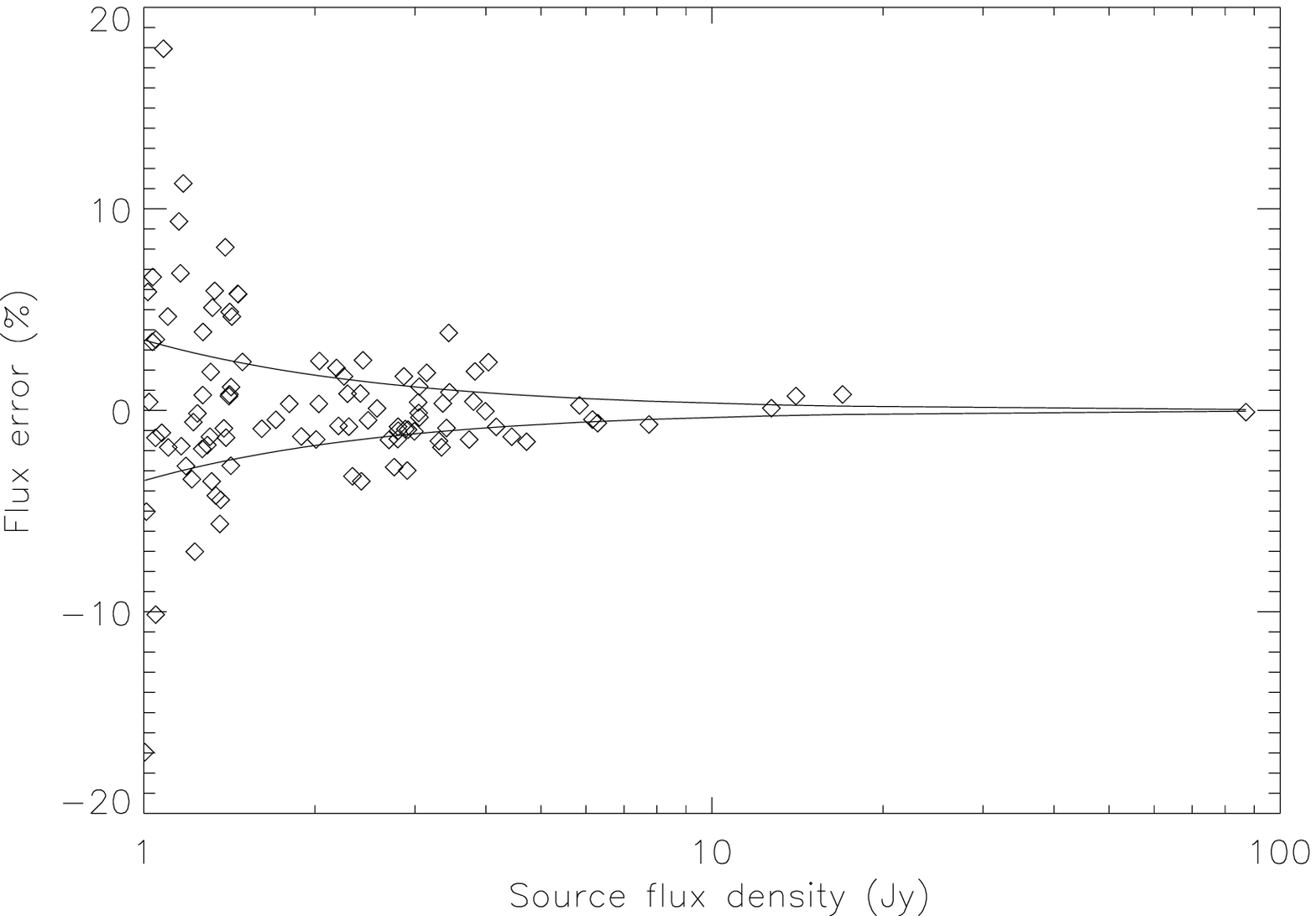}
\caption{Same as Figure~\ref{err_32T}, but for the 512T simulation. The synthesized beam is now $\Theta_b \sim 11.8$~arcmin. The solid line in the bottom figure indicates the SNR$^{-1}$ envelope.}
\label{err_512T}
\end{figure}

We computed the difference between the true flux density and position values and their final estimates (Figure~\ref{err_512T}), as was done for the 32T simulation.  

As in the 32T case, errors increase with decreasing flux densities and are well matched to their expected theoretical limits. The median and rms values show no systematic errors in the recovered parameters (Table~\ref{best_fit_512T}).
\begin{table}
 \centering 
  \caption{Median and rms values of the offset between the model and the fitted parameters for the simulated 512T image.}
  \begin{tabular}{@{}lcc@{}}
  Parameter	&	Median			&  	rms \\
  \hline
  RA		&	0.99~arcsec		&	17~arcsec	\\
  DEC		&	0.56~arcsec		&	11~arcsec	\\
  flux density	&	-0.15\%			&	4\%		\\
  \end{tabular}
 \label{best_fit_512T}
\end{table}

\begin{figure}
\centering
  \includegraphics[width=1.0\hsize]{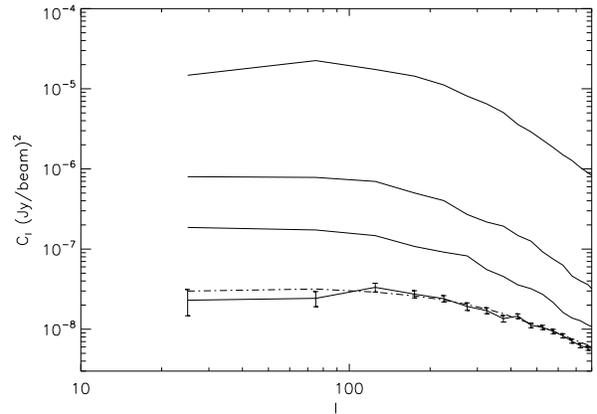}
\caption{Power spectra of residual images for an increasing number of subtracted sources. From top to bottom: initial image without subtracting any source (Figure~\ref{512T_original}), after subtracting fifteen sources (Figure~\ref{512T_residual1}), after subtracting 50 sources (Figure~\ref{512T_residual2}), after the whole sky model is subtracted (Figure~\ref{512T_residual3}). The error bars are at the 1$\sigma$ confidence level. The dashed line represents the noise power spectrum.}
\label{power_spectrum_512T}
\end{figure}

The power spectrum of the residual image after subtracting all 101 sources agrees with the expected thermal noise within 1$\sigma$ error for each multipole value, indicating that there is no significant statistical leftover from source subtraction (Figure~\ref{power_spectrum_512T}). The power spectra spans almost three orders of magnitude because the 512T simulation probes the $\log N$-$\log S$ at lower flux densities. This demonstrates that the algorithm is able to simultaneously remove a large number of sources which span two orders of magnitude in flux density.

This is a relevant result in the light of EoR measurements, where a high accuracy in source subtraction is required to achieve the necessary dynamic range. The detection of the EoR is believed to require this high accuracy in foreground subtraction because the cosmological signal is 5-6~orders of magnitude below the strongest sources in the sky. Due to the time constraints that come with the real time nature of the MWA, subtracting sources in the visibility domain - ``peeling'' - is only practical for the brightest sources which are also required to accurately constrain antenna primary beam models. In order to achieve accurate calibration and subtraction using these sources, a comprehensive global sky model is required. This will be obtained by surveying the sky in the first months of operation with the full array. At the same time, the actual tile beams will be measured and used to improve the beam models. The knowledge of the sky and the beams can be improved in a bootstrapping fashion by repeating the sky survey.

We expect that the initial $10^5-10^6$ dynamic range can be alleviated by 2-3~orders of magnitude through a very precise peeling procedure. A further subtraction of the remaining bright sources is required in the integrated images.

If the dynamic range is expressed as the ratio between the brightest source in the map and the noise rms, the source subtraction in the 512T case achieves a dynamic range of $\sim$3400 through our minimimization scheme.

It is also interesting to introduce the relative dynamic range, defined as the ratio between the true flux density of a source and the difference between the true and the recovered flux density (Pindor et al. 2010). This is another way of estimating the residual contamination due to an imperfect subtraction. Studies in the literature indicated that bright sources should be subtracted down to the 100-10~mJy level in order not to affect the subtraction of fainter foreground sources and, ultimately, the recovery of the EoR signal (Bowman et al. 2009, Liu et al. 2009b).

\begin{figure}
\centering
  \includegraphics[width=1.03\hsize]{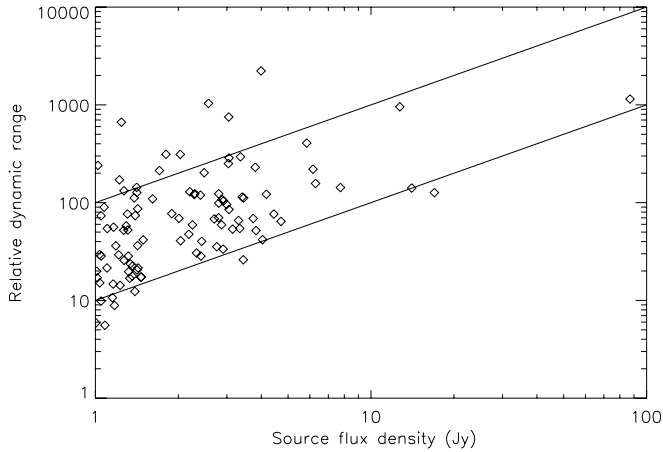}
\caption{The relative dynamic range as a function of flux density. The solid lines indicate the 10~mJy (upper) and the 100~mJy (lower) flux density threshold for source subtraction.}
\label{dynamic_range_512T}
\end{figure}

Figure~\ref{dynamic_range_512T} displays the relative dynamic range for source subtraction in the 512T simulation. It can be seen that, with the level of noise present in our simulated image, $\sim$92\% of the sources are above the 100~mJy threshold.  

Since we are fitting all the sources simultaneously and iteratively, the main limitation to the relative dynamic range comes from thermal noise rather than sidelobe contamination. Given the behaviour shown in Figure~\ref{err_512T}, we expect the dynamic range to increase if we consider a longer integration where the thermal noise decreases. Section~\ref{out_of_beam_sources} will confirm this statement.

\subsection{Out of beam sources}
\label{out_of_beam_sources}

An image has the limitation of excluding all the sources outside the image itself (out-of-beam sources). If visibility data were accessible, the information corresponding to out-of-beam sources would still be accessible and they could be subtracted in a traditional selfcalibration-deconvolution loop. Once the image is generated and visibility data discarded, information about out-of-beam sources is lost, apart from the sidelobes, which will still contaminate the image if they are bright enough.

We investigated how well our method subtracts out-of-beam sources by minimizing their sidelobe contribution to the image, i.e., by fitting the sidelobe pattern of a source, regardless of being able to image the source itself.

\begin{figure}
\centering
  \includegraphics[width=1.0\hsize]{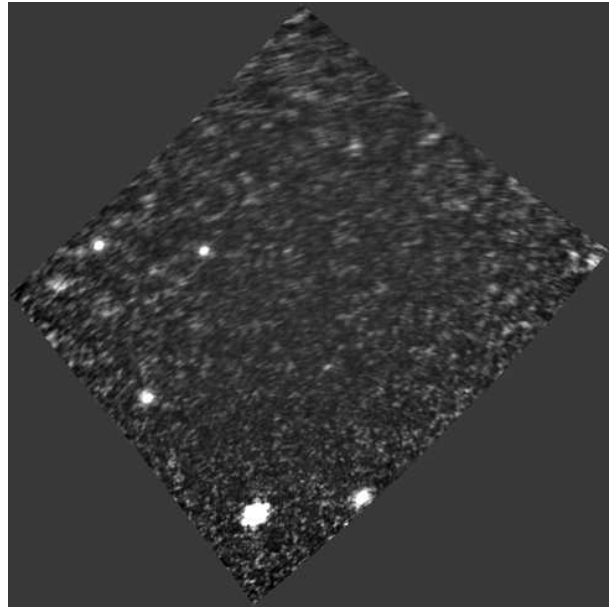}
\caption{Simulated image with the 10~min 512T $uv$ coverage. Five sources are within the field of view and two outside. The color scale runs between -0.3 and 1~Jy~beam$^{-1}$.}
\label{oob_original}
\end{figure}

In order to make sure that the out-of-beam source sidelobes have good SNR, we integrated individual 8~sec 512T snapshots up to 10~min in a 40~kHz channel (Figure~\ref{oob_original}). The thermal noise in our 10~min simulated image is 2.8~mJy~beam$^{-1}$.

In order to reduce the computational load, we included only the brightest seven sources used in the 512T simulation, therefore the faintest source is $\sim$6~Jy. Two sources were displaced from their previous position and moved two degrees outside the edge of the image. The out-of-beam sources had $\sim$14.1~Jy and $\sim$12.7~Jy flux densities respectively and we assumed that an initial estimate of their parameters is know from a pre-existing source catalogue. 

\begin{figure}
\centering
  \includegraphics[width=1.0\hsize]{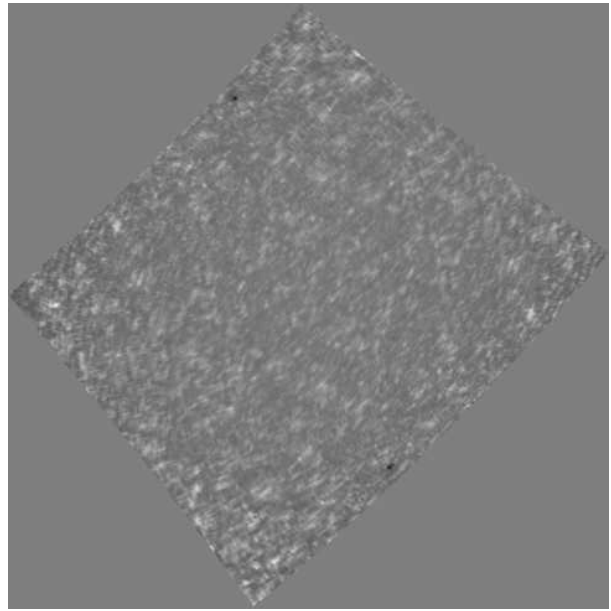}
\caption{Residual image with only the five in-beam sources subtracted, after five iterations. The color scale runs between -0.2 and 0.2~Jy~beam$^{-1}$. The five sources within the field of view were well removed revealing the sidelobe pattern of the unsubtracted out-of-beam sources.}
\label{oob_residual1}
\end{figure}

\begin{figure}
\centering
  \includegraphics[width=1.0\hsize]{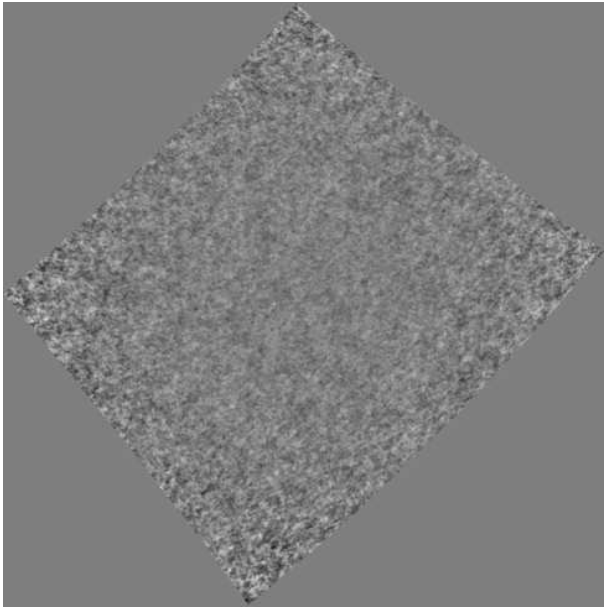}
\caption{Residual image where all the sources were subtracted. Five iterations were performed. The color scale runs between -0.03 and 0.03~Jy~beam$^{-1}$. The sidelobe pattern has been removed down to the thermal noise level.}
\label{oob_residual2}
\end{figure}

The subtraction was performed according to the following steps:
\begin{enumerate}  
  \item{} an initial parameter estimate of the five sources within the field of view was computed and the sources were subtracted ignoring the out-of-beam sources (Figure~\ref{oob_residual1});
   \item{} the two out-of-beam sources were included in the sky model and a joint parameter estimate performed. The best fit model is subtracted from the image (Figure~\ref{oob_residual2});
\end{enumerate}  

\begin{figure}
\centering
  \includegraphics[width=1.0\hsize]{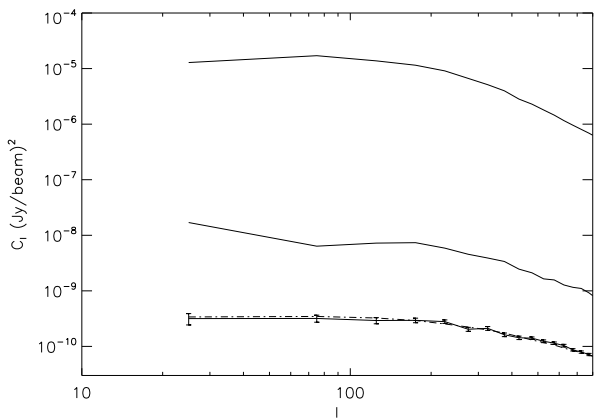}
\caption{Power spectra of residual images for an increasing number of subtracted sources. From top to bottom: initial image without subtracting any source (Figure~\ref{oob_original}), after the sources within the field of view were subtracted (Figure~\ref{oob_residual1}), after all the sources (in and out-of-beam) were subtracted (Figure~\ref{oob_residual2}). The error bars are at the 1$\sigma$ confidence level. The dashed line represents the noise power spectrum.}
\label{power_spectrum_512T_oob}
\end{figure}

The final image after the whole sky model was subtracted is consistent with the thermal noise level and its power spectrum agrees with the noise power spectrum at all angular scales (Figure~\ref{power_spectrum_512T_oob}). It is important to note that power spectrum of the residual image after removing only the sources within the field of view is still well above the expected noise power spectrum. The subtraction of the sidelobe pattern of the out-of-beam sources improves the dynamic range by a further factor of $\sim$5.
\begin{figure}
\centering
  \includegraphics[width=1.0\hsize]{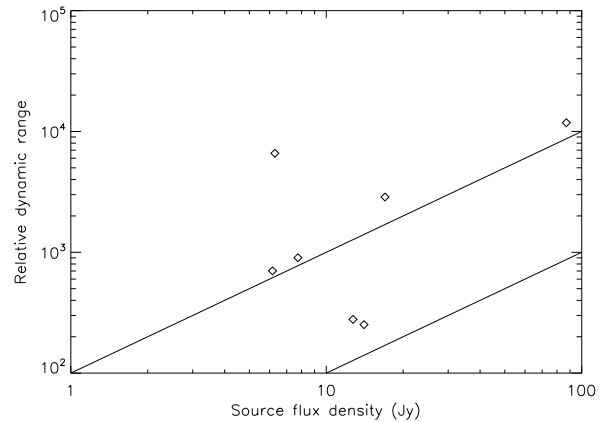}
\caption{As in Figure~\ref{dynamic_range_512T} but for the out-of-beam sources. The two sources with the poorest relative dynamic range are the out-of-beam sources for which the SNR is lower.}
\label{dynamic_range_512T_oob}
\end{figure}

The plot of the relative dynamic range (Figure~\ref{dynamic_range_512T_oob}) confirms the results of Section~\ref{512T_Results}. The relative dynamic range of the sources inside the field of view has improved by $\sim$2 orders of magnitude by longer integration and the two sources with the worst dynamic range are the out-of-beam sources, which have a poorer SNRs. All the sources within the field of view are now above the 10~mJy threshold.

{\subsection{Computational costs}
\label{computational_costs}

In our simulations we have shown that forward modeling can achieve a high level of precision in source subtraction. This comes, however, with a significant computational cost. Up to the number of sources that we have considered in our 512T simulation, the greatest computational load comes from imaging rather than generating the visibilities or fitting for the astrophysical parameters. In the case of an 8~sec snapshot image with a $20^\circ \times 20^\circ$ field of view we estimated that $\sim$70~Gflops are required to generate each image, therefore each iteration of the subtraction scheme requires $\sim$200~Gflops for every source that has to be subtracted. It takes $\sim$40~sec on a normal Dell 2.4~GHz 2~quad~Core machine.

Although such a computational need might require a very long processing time - particularly for long integrations -, there are several ways of shortening the processing time length. The most straightforward way is to implement a Graphics Processing Units (GPU) pipeline in the most computationally intensive part of the process, i.e. imaging. By running our simulations on a GPU, a factor $\sim$5 in time is gained. 

The second shortcut is to parallelize the forward modeling loop. Although the simulations presented in this work were performed in serial, the calculation of the forward model and the partial derivatives can be run in parallel, potentially on a dedicated GPU machine. 

Finally, it is important to notice that such a high level of precision in source removal might be superfluous for very faint sources for which calibration errors are larger. In this case, convenient approximations in fitting source positions (i.e, Pindor et al. 2010) will speed up the calculations and might eventually give the same level of accuracy in the subtraction.

\section{Conclusions}
\label{final_conclusions}

We have presented a point source deconvolution technique that makes use of forward modeling and an algebraic nonlinear minimization scheme. The main motivation for this implementation was achieving high dynamic range images in the absence of visibility data. Current (MWA) and future (SKA) radio interferometers require such a huge number of elements that they are being forced to rely more and more on real-time calibration and imaging, without the use of traditional selfcalibration techniques.

The basic idea of our scheme is to forward model the sky brightness, i.e., to filter the sky model through the same instrumental response that is applied to the data. In the case of radio point sources, the forward model is the synthesized beam which is generated for each source individually. In this way, position dependent variations of the synthesized beam are accounted for. 

Point source astrophysical parameters are recovered through a nonlinear minimization over the image pixels. In this way we overcome the known dynamic range limitations of image-based deconvolution due to pixelization effects. Since the presented technique minimizes all the sources simultaneously and in an iterative way, it is minimally sensitive to sidelobe noise and essentially limited by thermal noise.

It is worth noticing that this method can be applied to different sky components and can incorporate calibration parameters such as ionospheric displacements and primary beam shapes measured from the actual data.

The technique was applied to three different simulated cases: a 10~min integration with the 32T MWA, an 8~sec snapshot image of the 512T MWA, and a 10~min integration with the 512T MWA where sources were placed inside and outside the field of view. 

In all cases we were able to subtract sources down to the thermal noise without assuming an a priori knowledge of the sky, with the exception of initializing the position and flux density of sources placed outside the field of view.
The final residual images are consistent with the expected thermal noise on all the angular scales. Errors in the fitted parameters decrease with increasing SNR, in agreement with the expected theoretical measurement error distribution. Even when sources were not physically present in the images, we could subtract their sidelobes down to the thermal noise level.

The 512T simulations are relevant in the light of the MWA EoR experiment. Since only a limited number of sources can be subtracted in real time, an off-line subtraction of the residual sources will have to be performed on the images to a high level of accuracy in order to precisely remove them and their direction-dependent synthesized beams. 

In the simulation of an 8~sec image with the 512T array, we achieved a dynamic range of $\sim$3400, indicating that the subtraction of foreground sources can be improved by 3 orders of magnitude through this technique. Source parameters can be retrieved with an average error of 10~arcsec on positions and 0.15\% errors on flux densities.

The relative dynamic range of our subtraction is limited by the thermal noise and is above the 100~mJy threshold for 92\% of the sources. Since the best fit parameters improve with the SNR, a lower threshold - i.e. 10~mJy - can be reached by lowering the thermal noise through a longer integration. In fact, in the 512T 10~min simulation all the five sources present in the image had a dynamic range above the 10~mJy threshold, indicating that bright sources can be subtracted to a level that should not affect the detection of the EoR.

A sky model more realistic than only point sources could be forward modeled by modifying the procedure presented here. Extended sky emission modeled as a list of delta functions (i.e. the equivalent of CLEAN components) could be directly treated by the present approach. More sofisticated modeling of extended emission that uses a set of basis functions like, for instance, shapelets (Yatawatta 2010) or a principal component analysis (de Oliveira-Costa et al. 2008) can be incorporated by convolving the model of the brightness distribution with the instrumental primary beam and then sampling it according to the $uv$ distribution (see Wayth et al. 2010 for an example of this approach).

Future work will investigate these extensions and include a more realistic instrument model to better simulate the strategies for the EoR detection.

\section*{Acknowledgments}
We thank an anonymous referee for useful comments. This work was supported by the U.S. National Science Foundation under grants AST-0457585 and PHY-0835713.

\bibliographystyle{plain}

\label{lastpage}

\end{document}